\documentclass[reprint,amsmath,amssymb,superscriptaddress,aps,prl]{revtex4-2}
\usepackage{graphicx}
\usepackage{subfigure}
\usepackage{dcolumn}
\usepackage{tabularx}
\usepackage{bm}
\usepackage{braket}
\usepackage{gensymb}
\usepackage{color}
\usepackage{lineno}

\begin{document}
	\preprint{APS/123-QED}
	
	\title{Fault-Tolerant Information Processing with Quantum Weak Measurement}
	
	\author{Qi Song}
	\altaffiliation{These authors contributed equally to this work.}
	\affiliation{State Key Laboratory of Photonics and Communications, Institute for Quantum Sensing and Information Processing,  Shanghai Jiao Tong University, Shanghai 200240, People's Republic of China}
	
	\author{Hongjing Li}
	\email{lhjnet2012@sjtu.edu.cn}
	\altaffiliation{These authors contributed equally to this work.}
	\affiliation{State Key Laboratory of Photonics and Communications, Institute for Quantum Sensing and Information Processing, Shanghai Jiao Tong University, Shanghai 200240, People's Republic of China}
	\affiliation{Hefei National Laboratory, Hefei, 230088, People's Republic of China}
	\affiliation{Shanghai Research Center for Quantum Sciences, Shanghai, 201315, People's Republic of China}
	
	\author{Chengxi Yu}
	\affiliation{State Key Laboratory of Photonics and Communications, Institute for Quantum Sensing and Information Processing,  Shanghai Jiao Tong University, Shanghai 200240, People's Republic of China}
	
	\author{Jingzheng Huang}
	\affiliation{State Key Laboratory of Photonics and Communications, Institute for Quantum Sensing and Information Processing, Shanghai Jiao Tong University, Shanghai 200240, People's Republic of China}
	\affiliation{Hefei National Laboratory, Hefei, 230088, People's Republic of China}
	\affiliation{Shanghai Research Center for Quantum Sciences, Shanghai, 201315, People's Republic of China}

	\author{\\Ding Wang}
	\affiliation{State Key Laboratory of Photonics and Communications, Institute for Quantum Sensing and Information Processing, Shanghai Jiao Tong University, Shanghai 200240, People's Republic of China}
	
	\author{Peng Huang}
	\affiliation{State Key Laboratory of Photonics and Communications, Institute for Quantum Sensing and Information Processing, Shanghai Jiao Tong University, Shanghai 200240, People's Republic of China}
	\affiliation{Hefei National Laboratory, Hefei, 230088, People's Republic of China}
	\affiliation{Shanghai Research Center for Quantum Sciences, Shanghai, 201315, People's Republic of China}
	
	\author{Guihua Zeng}
	\email{ghzeng@sjtu.edu.cn}
	\affiliation{State Key Laboratory of Photonics and Communications, Institute for Quantum Sensing and Information Processing, Shanghai Jiao Tong University, Shanghai 200240, People's Republic of China}
	\affiliation{Hefei National Laboratory, Hefei, 230088, People's Republic of China}
	\affiliation{Shanghai Research Center for Quantum Sciences, Shanghai, 201315, People's Republic of China}
	
	\begin{abstract}	
		Noise is an important factor that influences the reliability of information acquisition, transmission, processing, and storage. Inspired by quantum weak measurement, we propose a fault-tolerant information processing approach to suppress inevitable noise effects. The approach employs pairwise orthogonal postselection bases with various small angles, and optimizes the composition of measurement outcomes as a decoding rule. Signals to be protected can be retrieved with minimal distortion after transmission through a noisy channel. Demonstrations using two-level superposition states and Einstein-Podolsky-Rosen states transmitted through decoherence noise channels show that the mean squared error distortion can approach zero, and fault-tolerant capability can reach unity with finite quantum resources. To experimentally verify the availability, classical coherent light and coherent state are used for encoding information. The approach potentially offers a solution for suppressing noise in long-distance quantum communication, high-sensitivity quantum sensing, and precise quantum computation. 
	\end{abstract}
	
	\maketitle
		\textit{Introduction---} Quantum information technologies refer to a superior way to acquire, transmit, process, and store information carried by quantum states, and have made great achievements in quantum computation \cite{divincenzo1995quantum,ofek2016extending,barz2013experimental,self2024protecting,doi:10.1126/science.abe8770,Ni2023beating,konno2024logical}, quantum sensing \cite{RevModPhys.89.035002,RevModPhys.90.035006,RevModPhys.90.040502,ligo2011gravitational}, quantum key distribution \cite{RevModPhys.94.035001,PhysRevLett.95.040503,fang2020implementation,PhysRevLett.91.087901,PhysRevLett.95.040503}, quantum teleportation \cite{PhysRevLett.70.1895,ma2012quantum,doi:10.1126/sciadv.adj3435,bouwmeester1997experimental,pirandola2015advances}, and quantum memory\cite{WOS:000272302800012,WOS:000304905300040,PhysRevLett.127.010503,Liu:20}, etc. However, the inevitable noises may degrade the coherence and entanglement of quantum states, greatly reducing the quantum enhancement performance, and seriously affecting the reliability of quantum information technologies in practical applications\cite{RevModPhys.88.041001,PhysRevLett.112.080801}.
		 
		Quantum error correction (QEC) provides a superior quantum way to decrease the noise effects on the information processing by introducing redundancy via specific coding methods to make multiple physical bits represent one logical bit, then the error detection and correction of logical bits can be realized\cite{PhysRevA.52.R2493, Kitaev1997Quantum, ofek2016extending, Ni2023beating}. Surface codes present a promising approach for large-scale quantum computing error correction\cite{WOS:001420297300001,Fowler86Surface,WOS:001595633200001}, and bosonic encoding has achieved notable advances—demonstrating logical qubit lifetimes that reach the break-even point\cite{ofek2016extending, Ni2023beating}. Nevertheless, practical deployment remains challenging due to platform-specific experimental constraints and substantial resource requirements.
	
		Photonic systems offer a compelling alternative as light provides the dominant quantum information carrier, such as that in quantum key distribution and quantum teleportation. Inspired by quantum weak measurement\cite{RevModPhys.86.307,XU2024Progress,Zhu2015weak,song2023}, this Letter proposes a fault-tolerant information processing (FTIP) protocol for photonic quantum states, including single photons, coherent states, and entangled photon pairs. Departing from QEC that protects quantum states, FTIP directly safeguards encoded information. According to the noise characteristics, pairwise orthogonal postselected measurement bases with tiny angles and optimal composition of measured results are chosen as the decoding rule. Then an arbitrary signal to be protected from noise effects can be retrieved with minimal mean squared error distortion by encoding on finite quantum states, while the fault-tolerant capability is significantly better than the classic information reliability protection ways in application field, noise suppression capability, and decoding latency. 
		
\begin{figure}[h!]
		\centering\includegraphics[width=8.6cm]{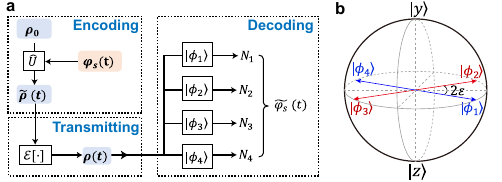}
		\caption{Schematic diagram of the FTIP approach. (a) an arbitrary signal to be protected is encoded onto the initial quantum states, then transmitted through a noisy channel, and finally decoded via postselected measurement bases. (b) Schematic diagram of four pairwise orthogonal measurement bases, $\varepsilon_{1,2}$ are symmetric with respect to a systematical axis $(\varepsilon_1+\varepsilon_2)/2$, and the angular between $\ket{\phi_1}$ and $\ket{\phi_2}$ is $2\varepsilon$.}	
		\label{fig1}
		\end{figure}

		\textit{Model Setup---} As depicted in Fig. \ref{fig1}a, the FTIP approach encompasses an encoding process, transmission through noisy channel, and a decoding process with postselected measurement bases. Let $\varphi_s(t)$ be an arbitrary signal to be protected from noise effects, and suppose $\rho_0$ be an arbitrary initial state, 
		\begin{equation}
			\rho_0=\sum_j p_j \ket{\psi_{0j}}\bra{\psi_{0j}},
		\end{equation}
		where $p_j$ is the probability of state $\rho_{0j}=\ket{\psi_{0j}}\bra{\psi_{0j}}$, and $\ket{\psi_{0j}}=cos\frac{\theta_j}{2}\ket{y}+sin\frac{\theta_j}{2}\ket{z}(0<\theta_j<\pi)$. Here, $\ket{y}$ and $\ket{z}$ may be physical state or logical state, and the variables $y$ and $z$ may be discrete or continuous. In the encoding process, $\varphi_s(t)$ is encoded as relative phase of the initial state $\rho_0$ according to an encoding rule ${\bf E}$, i.e., ${\bf E}:\varphi_s(t) \rightarrow \tilde\rho(t)$. One gets
		\begin{equation}
		\tilde\rho(t)=\hat{U}[\varphi_s(t)] \rho_0\hat{U}^{\dagger}[\varphi_s(t)]=\sum_j p_j \ket{\tilde\psi_{j}(t)}\bra{\tilde\psi_{j}(t)},
		\end{equation}
		where $\ket{\tilde\psi_{j}(t)}=cos\frac{\theta_j}{2}\ket{y}+e^{i\varphi_s(t)}sin\frac{\theta_j}{2}\ket{z}$. Subsequently, the encoded state is transmitted through a noisy channel and evolves to $\rho(t)$. The evolution caused by the interaction between quantum system and environment can be expressed as
		\begin{equation}
		\label{e}
		\rho(t)=\mathcal{E}[\tilde\rho(t)]=\sum_{k} E_{k}\tilde\rho(t)E_{k}^{\dagger},
		\end{equation}
		where Kraus operators satisfy the completeness condition of $\sum_{k} E_{k}^{\dagger}E_{k}=\mathcal{I}$, with $\mathcal{I}$ denoting the identity matrix. For an arbitrary d-dimensional system, $1\leq k \leq d^2$ \cite{nielsen2010quantum}. 
	
	The decoding process involves two steps, and obeys a decoding rule ${\bf D}: \rho(t)\rightarrow \widetilde{\varphi_s}(t)$(See details in Appendix). Firstly, $\rho(t)$ is detected by the postselected measurement bases with various tiny postselection angles, i.e.,
	\begin{equation}
		\label{p}
		\ket{\phi_{l}}=\frac{1}{\sqrt{2}}(\ket{y}+e^{i\varepsilon_l}\ket{z}),
	\end{equation}
	where $\varepsilon_l$ is the tiny postselection angle of the $l^{th}$ measurement base. Notably, at least two pairs of postselected measurement bases are orthogonal. Figure \ref{fig1}b depicts the example of using four postselected measurement bases, where $\{\ket{\phi_{1}},\ket{\phi_{4}}\}$ and $\{\ket{\phi_{2}},\ket{\phi_{3}}\}$ are pairwise orthogonal, $\varepsilon_{1}$ and $\varepsilon_{2}$ are symmetric with respect to the systematical axis $(\varepsilon_1+\varepsilon_2)/2$. The detection results are given by
	\begin{equation}
		\label{pn}
		N_l(t)= n_l P_l(t)=n_l Tr[\rho(t)\ket{\phi_l}\bra{\phi_l}],
	\end{equation}
	where $P_l(t)$ and $n_l$ are respectively detection probabilities and detected photon number in the $l^{th}$ base, and the total photon number in the initial state $\rho_{0}$ is $N=\sum_l n_l$. Secondly, the measured results are employed to retrieve $\varphi_s(t)$ with decoding rule of ${\bf D}$. According to the characteristics of the involved channel noises, one can select an optimal value of the systematical axis $(\varepsilon_1+\varepsilon_2)/2$ to determine the postselection angles $\varepsilon_l$ of the postselected measurement bases, then the signal $\varphi_s(t)$ may be retrieved with minimal distortion according to the decoding rule. Without loss of generality, we here consider the case of $|\varphi_s(t)|\ll1$. Then the results $\widetilde{\varphi_s}(t)$ may be retrieved by
	\begin{equation}
		\label{phi}
		\widetilde{\varphi_s}(t)=\frac{\xi(t)}{cot\varepsilon}\approx\chi\varphi_s(t)+\Delta\varphi_s(t),
	\end{equation}
	where $\varepsilon=(\varepsilon_2-\varepsilon_1)/2\ll1$, $\xi(t)$ is a calculated result according to the decoding rule based on the measured results $N_l(t)$ in the postselection measurement bases (See details in optimal composition strategy of Appendix), $\chi$ reflects the noise effects of the noisy channel on the encoded quantum state, and $\Delta\varphi_s(t)$ is the derivation resulted by measurement operations, which may be formally denoted as 
	\begin{equation}
		\label{d0}
		\Delta\varphi_s(t)=f(\chi, \varepsilon, N, \Upsilon),
	\end{equation}
	where $\Upsilon=\sum_j p_j sin\theta_j$ is caused by the encoding operations. Clearly, the ultimate goal is to choose an optimal decoding rule to lead $\chi\rightarrow1$ and $\Delta\varphi_s(t)\rightarrow0$, thereby completely suppressing the noise effects from the noisy channel and measurement operations.

	The performance of FTIP approach can be characterized by the distortion of the retrieved signal and the fault-tolerant capability. The distortion between the retrieved signal and the original signal can be evaluated by MSE distortion, which is given by 
	\begin{equation}
		\label{dis}
		d_{MSE}=E[(\widetilde{\varphi_s}(t) - \varphi_s(t))^2],
	\end{equation}
	where $E[\cdot]$ is the operation of averaging operation, and it can reach $d_{MSE}\rightarrow0$ when $\chi\rightarrow1$ and $\Delta\varphi_s(t)\rightarrow0$. The FTIP approach is considered fault-tolerant if it satisfies practical requirements regarding MSE distortion.  Actually, an optimal composition strategy in the decoding rule could lead $\chi\rightarrow1$, so that the distortion is mainly introduced by measurement noises. According to the requirement on MSE distortion, one can set a detection decision threshold $\Gamma$ in the decoding (See details in detection decision strategy of Appendix), and the fault-tolerant capability can be evaluated by the confidence of $\Delta\varphi_s(t)<\sqrt\Gamma$, which is formally expressed as 
	\begin{equation}
	\label{ft}
	F_t= g(\Gamma, \Delta\varphi_s).
	\end{equation}
	By optimally selecting the detection decision threshold $\Gamma$, one can achieve $F_t \to 1$, indicating that the proposed approach can reliably decode the original signal even when the retrieved results with errors, thereby enhancing the reliability of information transmission.

	\textit{Encoding with two-level superposition state---} In this case, the initial state $\rho_0$ is in 2-dimension Hilbert space, and the channel noise may be characterized by a 2-dimensional Kraus operator, which can be expressed as a universal expression of a 2$\times$2 complex matrix, i.e.,
	\begin{equation}
	\label{ee}
	E_k=\left[\begin{array}{cc}
		a_k & b_k \\
		c_k & d_k
	\end{array}\right].
	\end{equation}
	Correspondingly, arbitrary channel noise in the evolved quantum state $\rho(t)$ can be fully described using 12 real parameters with completeness condition. Utilizing pairwise orthogonal postselected bases and composition strategy, the noise effects can be effectively captured by 8 real parameters, i.e., $\mathcal{A}_1=\sum_k \textrm{Re}[a_k^*c_k]$, $\mathcal{A}_2=\sum_k \textrm{Re}[b_k^*d_k]$, $\mathcal{A}_3=\sum_k \textrm{Im}[a_k^*c_k]$,$\mathcal{A}_4=\sum_k \textrm{Im}[b_k^*d_k]$,$\mathcal{B}_1=\sum_k \textrm{Re}[a_k^*d_k]$, $\mathcal{B}_2=\sum_k \textrm{Re}[b_k^*c_k]$, $\mathcal{C}_1=\sum_k \textrm{Im}[a_k^*d_k]$, $\mathcal{C}_2=\sum_k \textrm{Im}[b_k^*c_k]$ (See more details in Supplementary Note S1). Clearly, one can refer to the above noises parameters to select the pairwise orthogonal postselected bases, and design the decoding rule to guarantee the accuracy and fault tolerance of the encoded information.

	We consider the decoherence noises\cite{nielsen2010quantum}, where $\mathcal{A}_{1,2,3,4}=0$ and $\mathcal{C}_{1,2}=0$, and the noise effects are characterized by $\chi_1=\mathcal{B}_1-\mathcal{B}_2$ and $\chi_2=\mathcal{B}_1+\mathcal{B}_2$(See more details in Supplementary Note S1). A special case is the noise induced by phase damping, phase flip, amplitude damping, and depolarizing, in which $\mathcal{B}_2=0$. To retrieve the original signal $\varphi_s(t)$ in this situation, the optimal choices of the four postselected measurement bases are
	\begin{equation}
	\label{mb}
	\begin{aligned}
		\ket{\phi_{1}}&=\frac{1}{\sqrt{2}}(e^{-i\frac{\varepsilon}{2}}\ket{0}+ie^{i\frac{\varepsilon}{2}}\ket{1}),\\
		\ket{\phi_{2}}&=\frac{1}{\sqrt{2}}(e^{i\frac{\varepsilon}{2}}\ket{0}+ie^{-i\frac{\varepsilon}{2}}\ket{1}),\\
		\ket{\phi_{3}}&=\frac{1}{\sqrt{2}}(e^{i\frac{\varepsilon}{2}}\ket{0}-ie^{-i\frac{\varepsilon}{2}}\ket{1}),\\
		\ket{\phi_{4}}&=\frac{1}{\sqrt{2}}(e^{-i\frac{\varepsilon}{2}}\ket{0}-ie^{i\frac{\varepsilon}{2}}\ket{1}).
	\end{aligned}
	\end{equation}
	Notes that when the channel is influenced by bit flip noise or bit-phase flip noise, two new measurement bases, $\{\ket{\phi_{5}},\ket{\phi_{6}}\}$, should be added in Eq.(\ref{mb}) to retrieve the original signal. In term of Eqs.(\ref{p}-\ref{phi}), the optimal composition of above measurement bases could lead noise effects to be suppressed, i.e., $\chi=\chi_1/\chi_2=(\mathcal{B}_1-\mathcal{B}_2)/(\mathcal{B}_1+\mathcal{B}_2)\rightarrow1$, and one can retrieve the signal as $\widetilde{\varphi_s}(t)=\varphi_s(t)+\Delta\varphi_s(t)$. 

	To illustrate the impact of quantum resources on signal retrieval, we analyze two paradigmatic encoding schemes with distinct photon statistics. First, suppose that the photon number of initial state obeys Poisson distribution with a mean value of $N$, such as coherent state, the signal could be encoded onto the polarization of a coherent state, and the retrieval performance is characterized by
	\begin{equation}	
	\label{d1}
	\begin{aligned}
		&\Delta{\varphi_s}(t)\approx
		\frac{\sqrt2}{\sqrt{N} \Upsilon cos\varepsilon (\mathcal{B}_1+\mathcal{B}_2)},\\
		&d_{MSE}\approx
		\frac{2}{N E[\Upsilon cos\varepsilon (\mathcal{B}_1+\mathcal{B}_2)]^2},\\
		&F_t=\sum_{q=n_{d1}}^{\infty}\frac{e^{-N}N^{q}}{q!},
		n_{d1}=\frac{2}{\Gamma [\Upsilon cos\varepsilon (\mathcal{B}_1+\mathcal{B}_2)]^2}.
	\end{aligned}
	\end{equation}
	Despite inherent photon number fluctuations, finite quantum resources enable $d_{\text{MSE}}\rightarrow0$ and $F_t\rightarrow1$ through appropriate threshold selection, even with significant distortion, as illustrated by the numerical simulation in Fig. S1. Contrastingly, when encoding $\varphi_s(t)$ on $N$ copies of single photon states, the photon number per copy becomes deterministic but introduces detection uncertainty. Here, photons are distributed across $L$ paths via multinomial statistics with a mean of $N/L$ and a variance of $N(L-1)/L^2$. With $\ket{\phi_{l}}$ ($l=1,2,3,4$), we obtain
	\begin{equation}
	\label{d2}
	\begin{aligned}
		&\Delta \varphi_s(t)\approx\frac{3\sqrt2}{4\sqrt{N} \Upsilon cos\varepsilon (\mathcal{B}_1+\mathcal{B}_2)},\\
		&d_{MSE}=\frac{9}{8N E[\Upsilon cos\varepsilon (\mathcal{B}_1+\mathcal{B}_2)]^2},\\
		&F_t=\sum_{q=n_{d2}}^{\infty}C_n^{q}(\frac{1}{4})^{q}(\frac{3}{4})^{N-q}, n_{d2}=\frac{9}{8\Gamma [\Upsilon cos\varepsilon (\mathcal{B}_1+\mathcal{B}_2)]^2}. 
	\end{aligned}
	\end{equation}
	Similarly, the result that finite quantum resources also allow $d_{\text{MSE}}\rightarrow0$ and $F_t\rightarrow1$ also holds, as confirmed by the numerical simulation in Fig. S1.

\begin{figure}[h!]
	\centering\includegraphics[width=8.6cm]{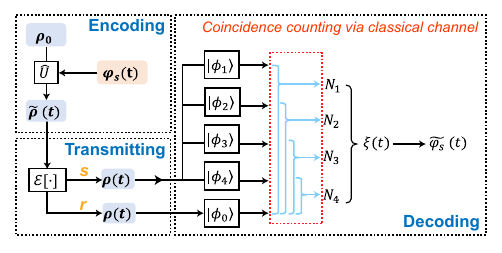}
	\caption{Schematic diagram of information encoding and decoding process of EPR state. After encoding original signal $\varphi_s(t)$ onto EPR state, the entangled photon pair is divided into two paths through the noisy channel. In data processing, the coincidence measurement is performed by the detection results transmitted through classic channel.}
	\label{fig3}		
	\end{figure}

	\textit{Encoding with EPR state---} Moreover, the FTIP approach can also be extended to scenarios involving encoding with entangled states. Taking the schematic diagram of encoding using the EPR state in Fig.\ref{fig3} as an example, encoding the original signal $\varphi_s(t)$ onto EPR state, $\rho_{0}=\sum_jp_j\ket{\psi_{0j}}\bra{\psi_{0j}}$ with $\ket{\psi_{0j}}=cos\frac{\theta_j}{2} \ket{H}_s\ket{H}_r + sin\frac{\theta_j}{2} \ket{V}_s\ket{V}_r$, the photon pairs are delivered into two paths through the noisy channel. In the detection, one needs at least to add a measurement base $\ket{\phi_{0}}=\frac{1}{\sqrt{2}}(\ket{H}_r+\ket{V}_r)$ for performing coincidence measurement, where the subscripts $s$ and $r$ indicate the two paths.

	The channel noises can be characterized by the 2-dimension Kraus operator \cite{WOS:000407374900015},
	\begin{equation}
	\rho (t)=\mathcal{E}[\tilde\rho(t)]=\sum_{k1,k2} (E_{k1} \otimes E_{k2}) \tilde\rho(t) (E_{k1}^{ \dagger} \otimes E_{k2}^{ \dagger}),
	\end{equation}
	and one gets $N_l$ by
	\begin{equation}
		N_l=Tr[\rho (t)\ket{\phi_{l}'}\bra{\phi_{l}'}],
	\end{equation}
	where $\ket{\phi_{l}'}=\ket{\phi_{0}}\otimes\ket{\phi_{l}}$. It is encouraging to note that the noise effects still reflect on the 8 real parameters in the multiplicative form in the FTIP approach, the noise effects can be characterized by $\chi_1=\mathcal{B}_1^2-\mathcal{B}_2^2$ and $\chi_2=(\mathcal{B}_1+\mathcal{B}_2)^2$ in decoherence noises channel.  The decoding rule $\bf D$ including postselected measurement bases $\ket{\phi_{l}'}=\ket{\phi_{0}}\otimes\ket{\phi_{l}} (l=1,2,3,4)$ leads $\chi=1$ under the noises induced by phase damping, phase flip, amplitude damping and depolarizing. By adding another two measurement bases $\ket {\phi_5'}$ and $\ket {\phi_6'}$, one may get $\chi\approx1$ when the encoded quantum state suffers bit flip and bit-phase flip (See Supplementary Note S2 for details). 

	Considering that EPR entangled photon pairs with $N$ copies are used and discretely detected by detectors, and  $\ket{\phi_{l}'}(l=1,2,3,4)$ are used. In this case, the original signal can be retrieved with
	\begin{equation}
	\begin{aligned}
		\label{d3}
		&\Delta \varphi_s(t)\approx\frac{3}{4\sqrt{N} \Upsilon cos\varepsilon (\mathcal{B}_1+\mathcal{B}_2)^2},\\
		&d_{MSE}\approx\frac{9}{16 N E[\Upsilon cos\varepsilon (\mathcal{B}_1+\mathcal{B}_2)^2]^2},\\
		&F_t=\sum_{q=n_{d3}}^{\infty}\frac{e^{-N}N^{q}}{q!}, n_{d3}=\frac{9}{16\Gamma [\Upsilon cos\varepsilon (\mathcal{B}_1+\mathcal{B}_2)^2]^2},
	\end{aligned}
	\end{equation}
	which yields $d_{\text{MSE}}\rightarrow0$ and $F_t\rightarrow1$ as well, as intuitively depicted in Fig. S2. 

\begin{figure}[h!]
		\centering
		\includegraphics[width=8.6cm]{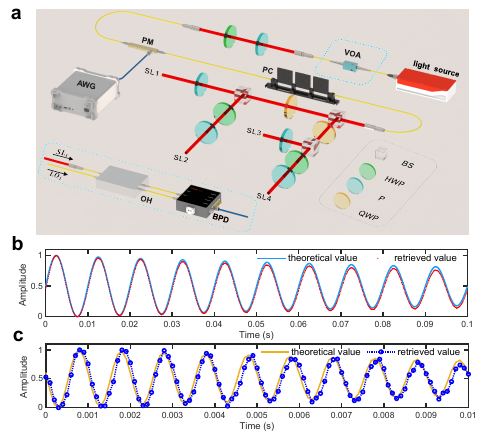}
		\caption{Experimental implementation of the FTIP approach. (a) Experiment setup. The implementation of using coherent state is enclosed by blue boxes, including the attenuation of coherent light via variable optical attenuator and homodyne detection on the signal light. The abbreviations in the figure are as following, VOA: variable optical attenuator, P: polarizer, FC: fiber collimator, PM: phase modulator, AWG: arbitrary waveform generator, PC: polarization controller, BS: beam splitter, HWP: half-wave plate, QWP: quarter-wave plate. SL: signal light, LO: local oscillator light, OH: free space passive optical hybrid, BPD: balanced photodetector. (b) The retrieved results when using classical coherent light. (c)The retrieved results when using coherent state. }	
		\label{fig2}
	\end{figure}

	\textit{Experimental verification---} To demonstrate feasibility, we simulate an information transmission scenario with the experimental setup shown in Fig. \ref{fig2}a.	Confirmatory experiments are presented via coherent light and coherent state as typical examples. Classical information is encoded onto the initial polarization state of via a phase modulator, followed by simulated decoherence noises using a rotating half-wave plate. The information is then recovered using the fault-tolerant measurement and decoding strategy(See Supplementary Note S3 for details). 

	The retrieved results using classical light in Fig. \ref{fig2}b closely match theoretical values, with only minor distortions from additional phase noise inevitably introduced when rotating the half-wave plate stochastically and electric noises. We further test with coherent state, which is generated by attenuating coherent light at approximately -75 dBm. Despite the severe attenuation, FTIP maintains robust recovery performance, as shown in Fig. \ref{fig2}c. Although the low rotation speed of the half-wave plate limits the experimental signal frequency, the phase modulator can achieve modulation rates up to tens of GHz, and high-speed acquisition devices enable rapid detection. Thus, FTIP is well-suited for high-frequency signals and rapidly varying channels, meeting the demands of high-speed communication.

	\textit{Discussion---} To suppress the noise effects resulted from interaction between quantum system and environment, we propose a FTIP approach via quantum weak measurement for photonics platform, which possesses two significant advantages. On one hand, an optimal composition strategy can achieve $\chi \rightarrow 1$ under decoherence noises, thereby completely suppressing the noise effects. On the other hand, there exists redundancy in the postselected measurement bases within the decoding rule. If one of the postselected measurement bases fails, the FTIP approach can still decode the information, significantly enhancing its robustness. Potentially, the approach may offer a viable solution for mitigating noise effects in long-distance quantum communication, high-sensitivity quantum sensing, and precise quantum computation. 	

	As we demonstrated in experiment, the FTIP approach is also suitable for classical light as initial state. The signal can be encoded on the polarization of the light field, and the parameters including $\Delta\varphi_s$, $d_{MSE}$ and $F_t$ can be obtained via particular distribution of light intensity, for example, when classic coherent light is used, one may also retrieve the original signal with Eq. (\ref{d1}), which means the proposed approach is available for quantum communication as well as classic communication.  Compared to classical information reliability protection methods\cite{proakis2008digital}, the FTIP approach exhibits the following superior performance. Firstly, classical reliability protection methods are primarily used for digital signal transmission to ensure reliability, whereas the FTIP approach can protect both analog and digital signals. Secondly, in classical optical communication, signals are typically encoded on the amplitude and phase of the light field, making them susceptible to amplitude noise and phase noise\cite{7174950}. Different from that, the FTIP approach encodes the signal on the relative phase between $\ket{y}$ and $\ket{z}$, which is immune to the phase noise on the overall phase of the light field. With the decoding rule $\bf{D}$, photon number fluctuation caused by amplitude noise of light field exerts little influence on the retrieved result of $\varphi_s(t)+\Delta\varphi_s(t)$, only causing the change of $\Delta\varphi_s(t)$. Finally, the FTIP approach has a low decoding complexity and short decoding latency as the information can be retrieved via direct measurement and simple arithmetic, which further provides the solution for real-time decoding.

\textit{Acknowledgments---} This work was supported by the National Natural Science Foundation of China (No.62471289), Natural Science Foundation of Shanghai (No.24ZR1432900), Quantum Science and TechnologyNational Science and Technology Major Project  (No.2021ZD0300703) and Shanghai Municipal Science and Technology Major Project (No.2019SHZDZX01).

\bibliography{prlref}	

~\\
\textit{Appendix: The decoding rule---} The decoding rule $\bf {D}$ includes two parts, i.e., optimal composition strategy and the detection decision strategy. In the optimal composition strategy, the aim is to construct an optimal composition with the measurement results from the postselected  measurement bases so that one can obtain Eq.(\ref{phi}). Here, we take four postselected measurement bases as an example, by choosing arbitrary three of four detection results $N_l(t)$ with the following way, one gets
\begin{equation}
	\label{xi}
	\begin{aligned}
		\xi(t)&=\frac{N_1(t)-N_3(t)}{N_2(t)-N_1(t)}\\&=K h(\chi,\varphi_s(t),\Delta \varphi_s(t))\\&\approx K[\chi\varphi_s(t)+\Delta \varphi_s(t)],
	\end{aligned}
\end{equation}
where $K=sin\frac{\varepsilon_1-\varepsilon_3}{2}/sin\frac{\varepsilon_2-\varepsilon_1}{2}$ is the amplification effect factor. When $\varepsilon\ll1$ and $\varepsilon_2-\varepsilon_3=\pi$ ($\ket{\phi_3}$ is orthometric to $\ket{\phi_2}$), one may get $K\gg1$ which corresponds to a clearly amplification effect (See Supplementary Note S1 for details). With the processing of Eq.(\ref{xi}), the noise effects on the encoded state $\chi=\chi_1/\chi_2$ can be divided into two parts, $\chi_1$, and $\chi_2$ respectively denote the noise effects on $N_1(t)-N_3(t)$ and $N_2(t)-N_1(t)$. 

In order to collect all results from the initial state $\rho_0$, measurement results from the base $\ket{\phi_4}$, which is orthometric to $\ket{\phi_1}$, are collected. Subsequently, one may obtain $\xi(t)$ in the following ways, i.e.,
\begin{equation}
	\begin{aligned}
		&\xi_1(t)=\frac{N_1(t)-N_3(t)}{N_2(t)-N_1(t)},
		\xi_2(t)=\frac{N_2(t)-N_4(t)}{N_2(t)-N_1(t)},\\
		&\xi_3(t)=\frac{N_1(t)-N_3(t)}{N_4(t)-N_3(t)},
		\xi_4(t)=\frac{N_2(t)-N_4(t)}{N_4(t)-N_3(t)}.
	\end{aligned}
\end{equation}
Simple calculation gives
\begin{equation}
	\label{md1}
	\xi_{1,2,3,4}(t)=cot\varepsilon[\chi\varphi_s(t)+\Delta \varphi_s(t)]
\end{equation}
and the amplification factor is obtained by $K=cot\varepsilon$. According to the noise characteristic, one can choose the value of $\varepsilon_2+\varepsilon_1$ or introduce additional pairwise orthogonal measurement bases to reach the aim of $\chi\rightarrow1$ (See Supplementary Note S1 for details).

In terms of the error transfer theory and error synthesis, one gets	
\begin{equation}
	\label{md2}
	\Delta^2 \xi= \frac{\Delta^2N_1+\Delta^2N_3}{(N_2-N_1)^2}+\frac{(\Delta^2N_2+\Delta^2N_1)(N_1-N_3)^2}{(N_2-N_1)^4},
\end{equation}
where $N_l\approx\frac{N}{4}$ and $\Delta N_{l}$ is the uncertainty of $N_l$. Generally, $\Delta N_{l}$ is associated with the distribution of initial quantum state $\rho_0$. Then one may get 
\begin{equation}
	\label{md3}
	\Delta{\varphi_s}(t)= \frac{\Delta \xi}{cot \varepsilon}.
\end{equation}
Combining with Eqs.(\ref{md1}-\ref{md3}), $\Delta{\varphi_s}(t)$ can be expressed as Eq.(\ref{d0}).

In the detection decision strategy, the aim is to determine the confidence so that an acceptable and optimal threshold of the retrieved phase may be chosen. By dividing the phase of a continuous value into $M-1$ intervals, $M$ is an odd number, $\varphi_s$ can be expressed as $(m-\frac{M+1}{2})\sqrt\Gamma$, $m=1, 2, 3, \cdots, M$, where $\Gamma$ is the detection decision threshold. If the uncertainty $\Delta{\varphi_s}(t)$ of the retrieved signal could be accepted by the detection decision threshold, i.e., 
\begin{equation}
	\label{dc}
	\Delta{\varphi_s}(t)\leq\sqrt\Gamma,
\end{equation}
the retrieved signal may be regarded as correctness. Then the fault-tolerant capability can be expressed as
\begin{equation}
	F_t=Pr\{\Delta{\varphi_s}(t)\leq\sqrt\Gamma\},
\end{equation}
where $Pr\{\Delta{\varphi_s}(t)\leq\sqrt\Gamma\}$ is the correctness confidence, associated with the distribution of initial state $\rho_0$. In this case, one can get $F_t \rightarrow 1$. 

Combining with the linear extraction of $\varphi_s(t)$, i.e., $|\varphi_s(t)|\ll1$, one gets 
\begin{equation}
	\Gamma\ll(\frac{2}{M+1})^2,
\end{equation}
which indicates that a smaller $\Gamma$ may allow a bigger $M$, and $\varphi_s$ can carry more information via a finite quantum states or a quantum state with finite particle number. The corresponding optimal threshold is given by 
\begin{equation}
	(\Delta{\varphi_s}(t))^2\leq\Gamma\ll(\frac{2}{M+1})^2,
\end{equation}
which means that one may choose an optimal detection decision threshold based on the quantum resource employed in the initial state.
\end{document}